\documentclass[10pt,prl,aps,twocolumn,superscriptaddress,amsfonts,amsmath,amssymb,floatfix]{revtex4-2}

\usepackage[utf8]{inputenc}
\usepackage[english]{babel}
\usepackage{upgreek}
\usepackage{amsmath,amsfonts,amsthm,amsbsy,empheq}
\usepackage{mathtools}
\usepackage[caption=false]{subfig}
\usepackage{graphicx,bm}
\usepackage{physics}
\usepackage{dsfont}
\usepackage{xcolor}
\usepackage{breqn}
\usepackage{ulem}
\usepackage{polski}
\usepackage{url}
\usepackage{newtxtext}

\DeclareTextCompositeCommand{\k}{LY1}{e}
{\oalign{e\crcr\noalign{\kern-.27ex}\hidewidth\char7\hidewidth}}

\colorlet{linkequation}{red}
\usepackage[colorlinks,linkcolor=red]{hyperref}

\newcommand{\vare}{\varepsilon}

\newcommand{\rmi}{{\rm i}}

\usepackage{xcolor}

\makeatletter
\let\cat@comma@active\@empty
\makeatother

\newcommand{\eqcontrib}{These two authors contributed equally to this work.}

\def \FigOne{
\begin{figure}[t]
\centering
\includegraphics[width=80mm]{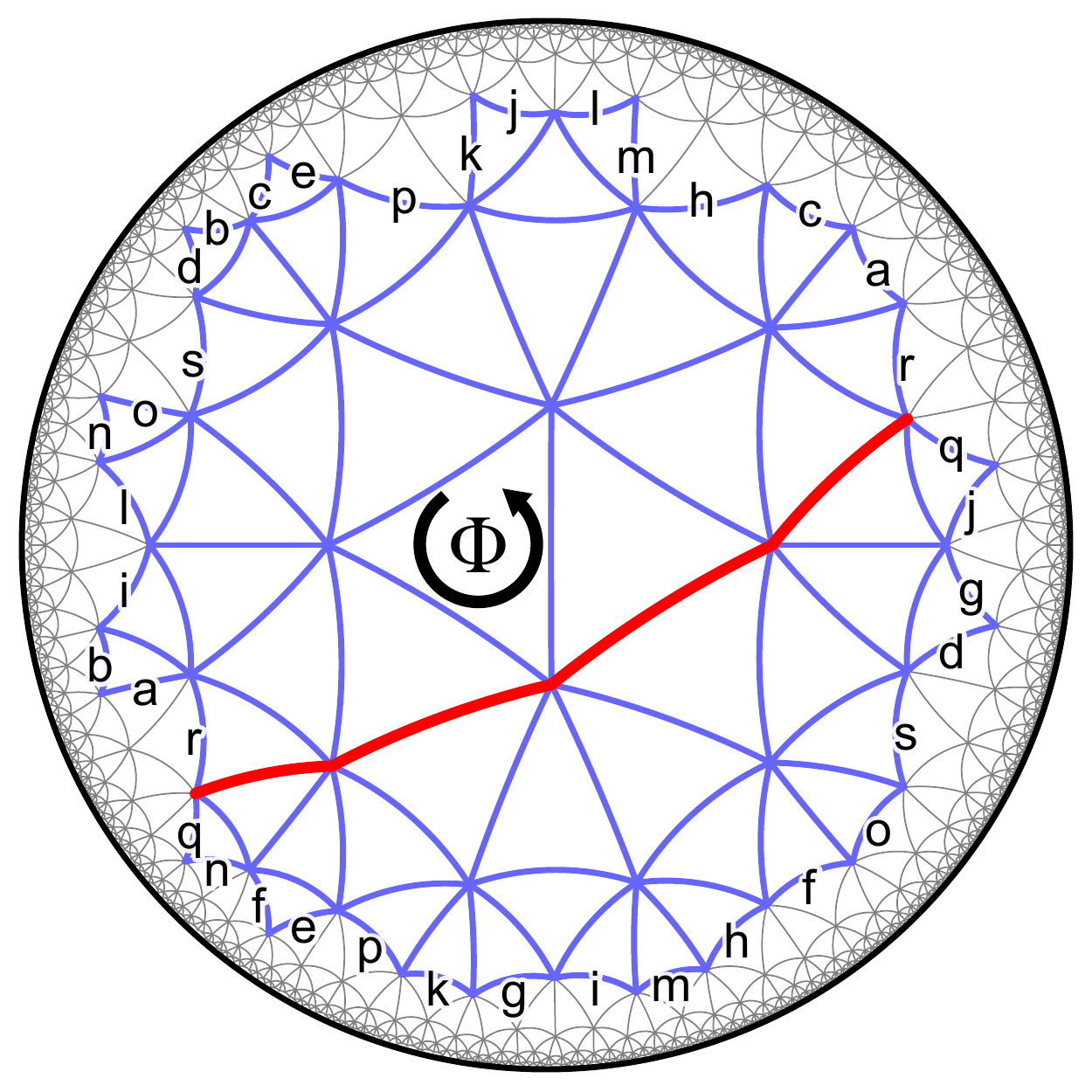}
\caption{We compute the Hofstadter butterfly spectrum on finite hyperbolic lattices with periodic boundary conditions to eliminate boundary effects. The example shows a $\{3,7\}$ graph with 24 vertices (blue) in the Poincar\'{e} disk that, when identifying the edges labelled by the same letter, yields a tessellation of a closed, genus-3 hyperbolic surface by triangles with coordination number 7 \cite{Boettcher2021}. The constant magnetic field is implemented by threading a flux $\Phi$ through each plaquette (shown here for one plaquette). Compactified finite hyperbolic $\{p,q\}$ lattices generally lead to tessellations of closed Riemann surfaces with genus $g\geq 2$, characterized by $g$ holes and $2g$ incontractible Aharonov--Bohm loops (one shown in red).}
\label{Fig1}
\end{figure}
}

\def \FigTwo{
\begin{figure*}[t]
\centering
\includegraphics[width=183mm]{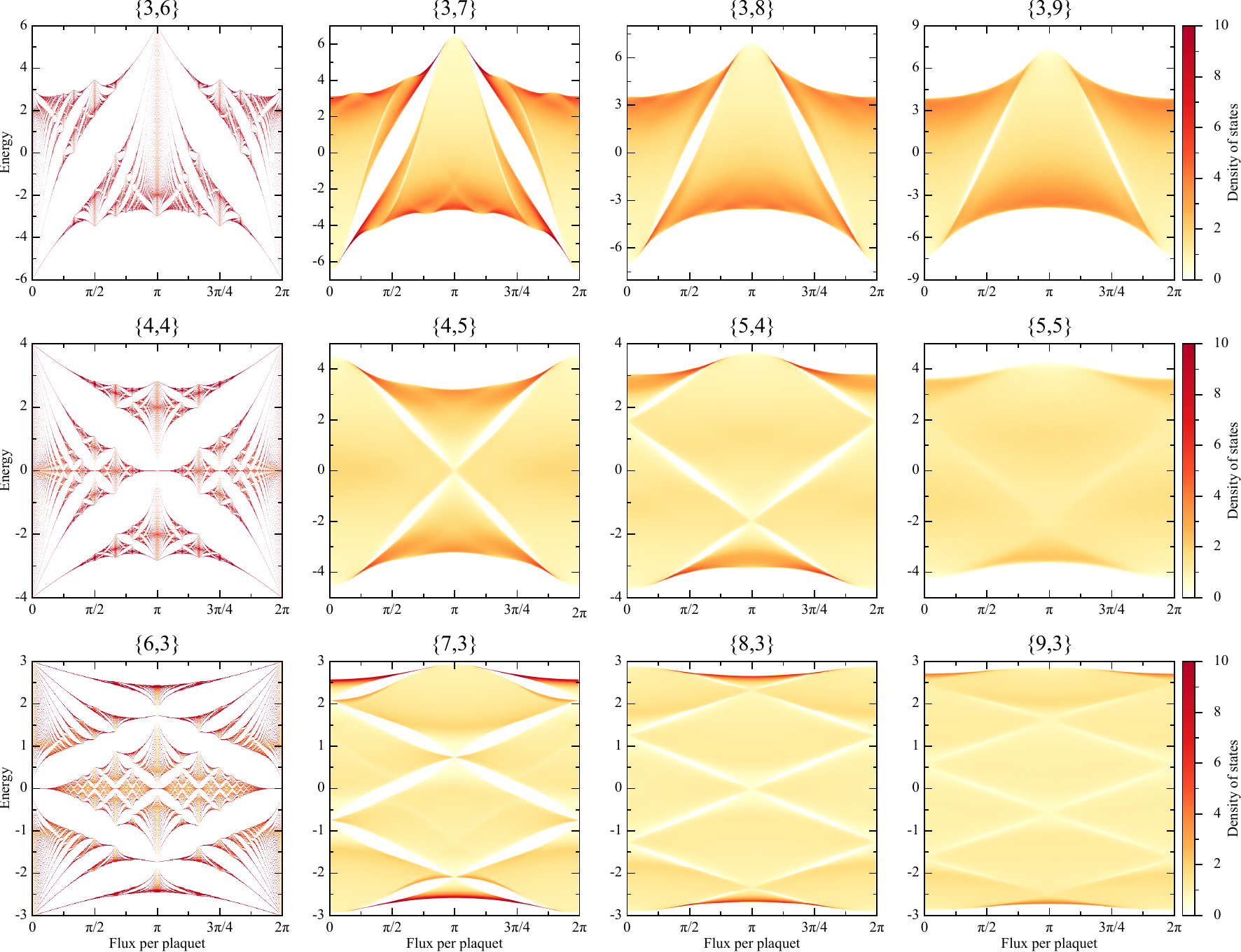}
\caption{Selection of Hofstadter butterfly spectra on $\{p,q\}$ lattices made from $p$-gons with coordination number $q$. The first column shows the Euclidean triangular, square, and hexagonal lattices, while the remaining columns correspond to genuine hyperbolic tessellations with $(p-2)(q-2)>4$. Upon varying $q$ for fixed $p$ in the hyperbolic case, we identify a striking universality: the shape of the butterfly, i.e., the number of large extended gapped regions, is solely determined by $p$, while $q$ has only a mild quantitative influence. No fractal structure exists on hyperbolic lattices. The figures show the normalized density of states over energy and magnetic flux per plaquette. To determine the hyperbolic spectra, we used graphs with periodic boundary conditions and up to several thousand vertices, corresponding to genera $g\sim 100$, and a sampling over Aharonov--Bohm fluxes \cite{SOM}. The infinite hyperbolic lattice would be obtained in the limit $g\to \infty$.}
\label{Fig2}
\end{figure*}
}

\def \FigThree{
	\begin{figure}[t]
		\centering
		\includegraphics[width=86mm]{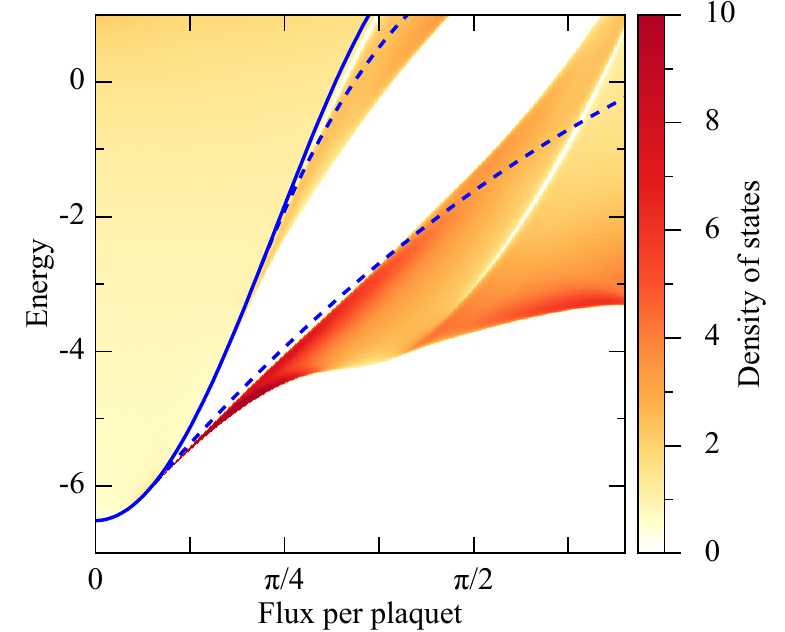}
\caption{Comparison of the HB spectrum on the  $\{3,7\}$ lattice with the energy spectrum of the continuum model in the long-wavelength limit. The continuum theory features both discrete Landau levels (dashed blue lines) and a continuous spectrum (bound from below by the solid blue line). Landau levels appear only for sufficiently large flux, which partially explains the absence of fractality of the HB spectrum on hyperbolic lattices. The lowest Landau level (lower curve) is clearly visible and develops into the low-energy part of the HB spectrum as a function of $\Phi$, separated by an extended gapped region from the rest of the spectrum. At the top of the figure, we observe a narrow region of reduced density of states to separate the second Landau level from the continuum.}
		\label{Fig3}
	\end{figure}
}

\def \FigureSOne{
	\begin{figure}[t]
		\centering
		\includegraphics[width=145mm]{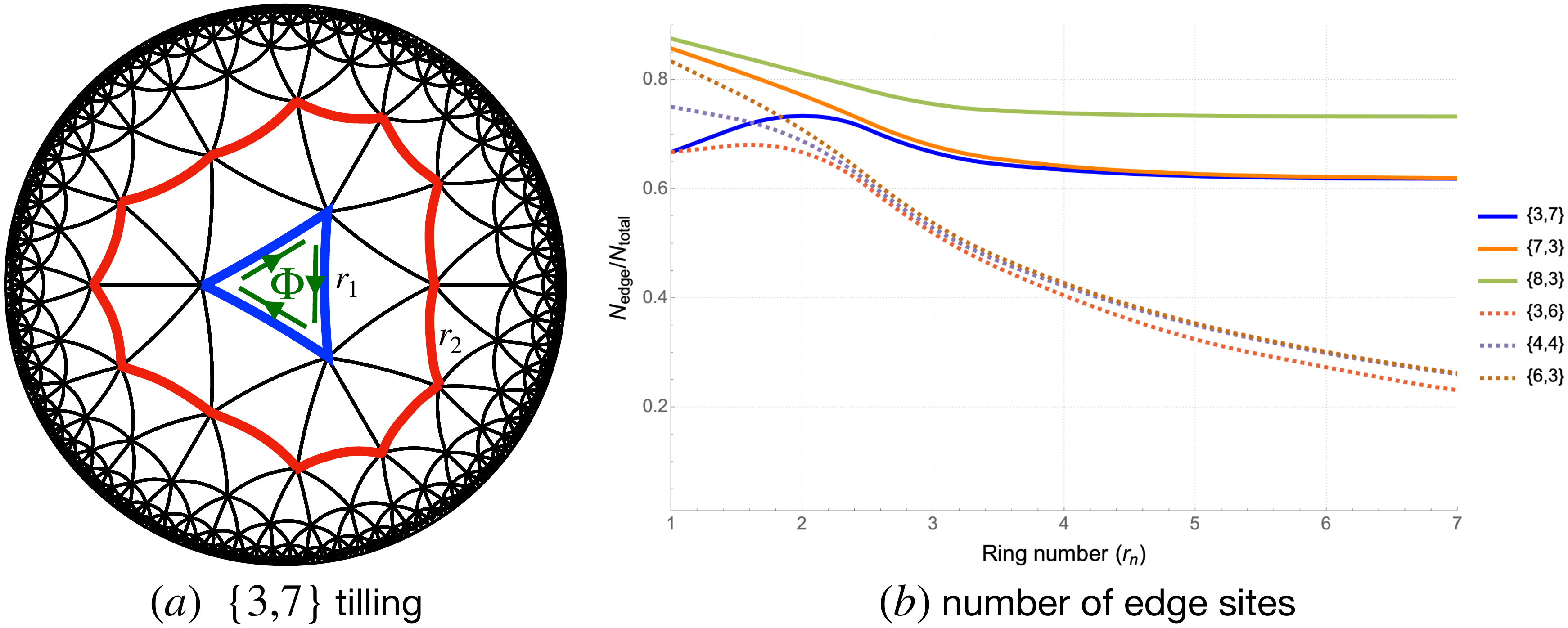}
		\caption{$ (a) $ Open boundary realization of a hyperbolic lattice with a ring-like geometry, where $ r_{n} $ denotes the $ n- $th ring that is the boundary of termination. A constant magnetic field through this finite lattice is obtained by threading a magnetic flux $\Phi$ through each plaquette. $ (b) $ The number of boundary sites compared to the total number of sites as $n$ increases saturates at a significant value in the hyperbolic case (thick curves). In the Euclidean case (dashed curves), as $n$ increases, the fraction of boundary sites decreases and vanishes asymptotically.}
		\label{SFig1}
	\end{figure}
}

\def \FigureSTwo{
	\begin{figure}[htb]
		\centering
		\includegraphics[width=184mm]{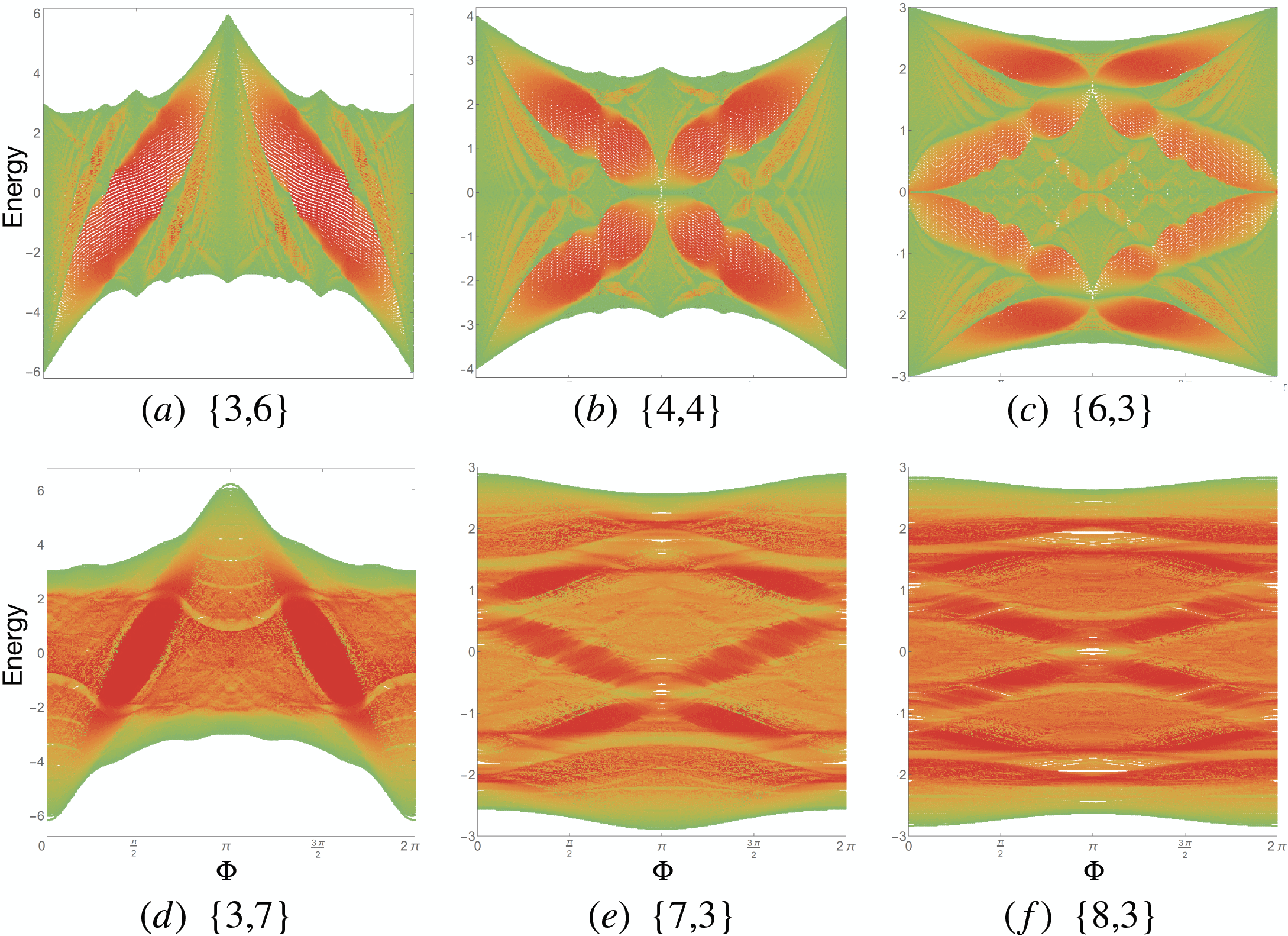}
		\caption{Hofstadter butterfly spectra for graphs with open boundary conditions, Euclidean (top row) and hyperbolic (bottom row). The color in the plot designates the contribution from bulk states (green) and the boundary states (red). The other color (orange) reflects energies from  mostly boundary states and some bulk states at the same energy. In the case of Euclidean lattices, the contribution from the boundary states is reduced in comparison to the hyperbolic spectra. The number of rings (vertices) used to compute the individual spectra is: for the Euclidean cases, $ (a) $ 22 (1387), $ (b) $ 19 (1369), $ (c) $ 16 (1441), and for the hyperbolic cases $ (d) $ 7 (1625), $ (e) $ 6 (1321), $ (f) $ 5 (1369).}
		\label{SFig2}
	\end{figure}
}

\def \FigureSThree{
	\begin{figure}[b]
		\centering
		\includegraphics[width=184mm]{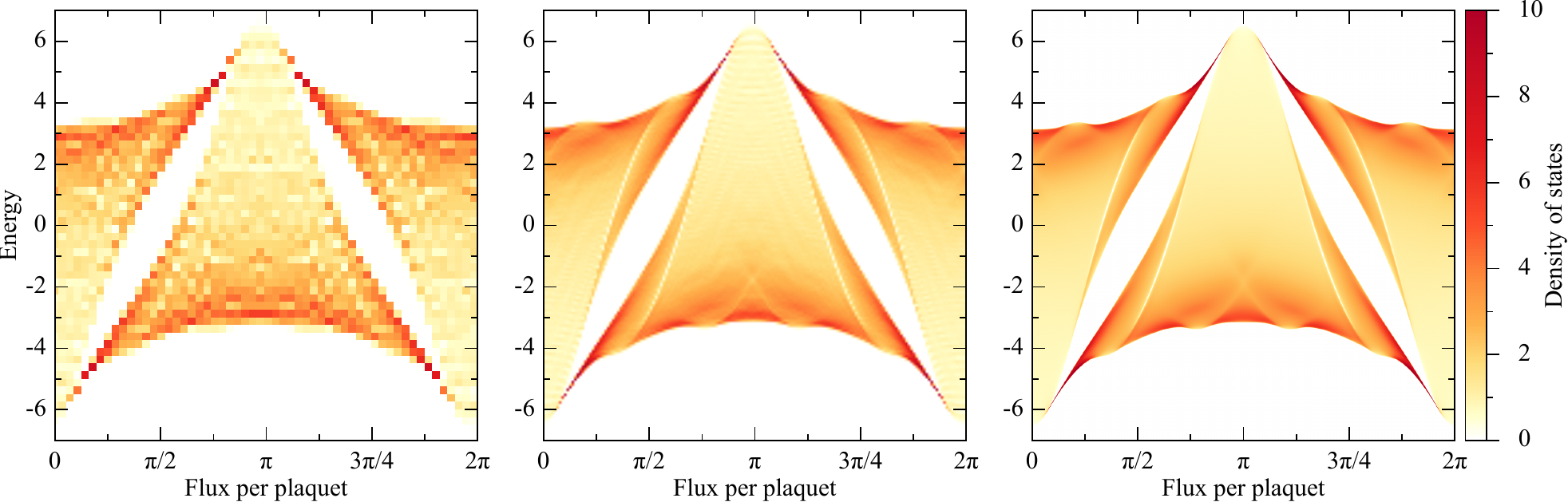}
		\caption{Comparison of the calculated Hofstadter butterfly for a genus 3 (left), genus 7 (middle) and genus 129 (right) $\{3,7\}$ regular map.}
		\label{Fig:A3}
		\vspace*{3mm}
		\includegraphics[width=184mm]{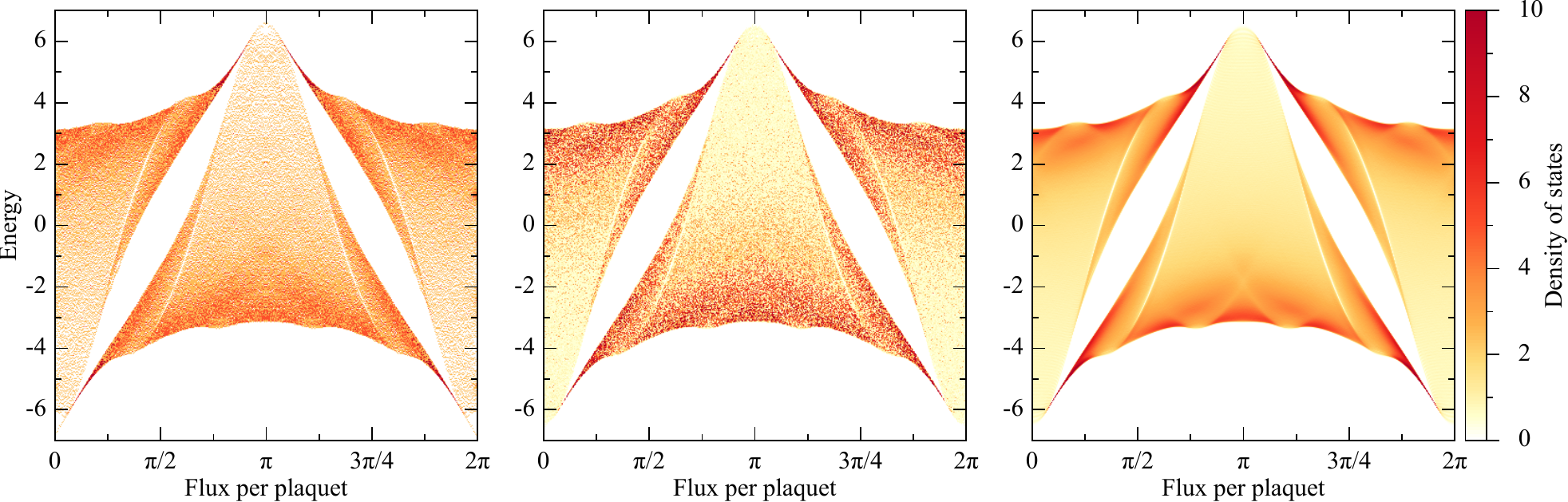}
		\caption{Comparison of the calculated Hofstadter butterfly for a genus 17 $\{3,7\}$ regular map. For the left plot, only one Aharonov--Bohm flux configuration was sampled, for the middle one ten, and for the right one 1000 flux configurations.}
		\label{Fig:A4}
	\end{figure}
}

\begin{document}

\hypersetup{pdftitle={title}}
\title{Universality of Hofstadter butterflies on hyperbolic lattices}

\author{Alexander Stegmaier}\thanks{\eqcontrib{}}
\affiliation{Institut f\"{u}r Theoretische Physik und Astrophysik, Universit\"{a}t W\"{u}rzburg, 97074 W\"{u}rzburg, Germany}

\author{Lavi K. Upreti}\thanks{\eqcontrib{}}
\affiliation{Institut f\"{u}r Theoretische Physik und Astrophysik, Universit\"{a}t W\"{u}rzburg, 97074 W\"{u}rzburg, Germany}

\author{Ronny Thomale}
\affiliation{Institut f\"{u}r Theoretische Physik und Astrophysik, Universit\"{a}t W\"{u}rzburg, 97074 W\"{u}rzburg, Germany}

\author{Igor Boettcher}
\affiliation{Department of Physics, University of Alberta, Edmonton, Alberta T6G 2E1, Canada}
\affiliation{Theoretical Physics Institute, University of Alberta, Edmonton, Alberta T6G 2E1, Canada}

\date{\today}

\begin{abstract}
Motivated by recent realizations of hyperbolic lattices in superconducting waveguides and electric circuits, we compute the Hofstadter butterfly on regular hyperbolic tilings. Utilizing large hyperbolic lattices with periodic boundary conditions, we obtain the true bulk spectrum unaffected by boundary states. The butterfly spectrum with large extended gapped regions prevails, and its shape is universally determined by the fundamental tile, while the fractal structure is lost. We explain how these features originate from Landau levels in hyperbolic space and can be verified experimentally.
\end{abstract}

\maketitle

Culminating in the quantum Hall effect and its descendants, the motion of charged quantum particles in a magnetic field forms the cradle of topological matter. The classical circular orbits of electrons are characterized by the cyclotron frequency $\omega_B = eB/m$, where $e$ and $m$ are the charge and mass of the electron, respectively, and $B$ is the magnetic field strength. 
In quantum mechanics, this circular motion is reflected by the formation of discrete Landau levels, which are highly degenerate eigenenergies that replace the continuous spectrum of free electrons. As we confine the motion to a lattice of lattice constant $a$, a fundamentally new physical regime appears when the magnetic length $\ell_B=\sqrt{\hbar/eB}$, expressing the quantum uncertainty in the position of the particle, becomes of the order of $a$. In this case, the energy spectrum of a single electron as a function of the magnetic flux per pla\-quette, $\Phi=Ba^2$, shows a fractal structure as $\Phi$ is increased from zero to the flux quantum $\Phi_0=2\pi\hbar/e$: for rational values $\Phi/\Phi_0=P/Q$ with integers $P,Q$, the spectrum consists of $Q$ bands, whereas for nonrational values of $\Phi/\Phi_0$
the spectrum is an infinite Cantor set. Furthermore, the spectrum features extended gapped regions without states, which form the wings of the celebrated Hofstadter butterfly (HB) \cite{Hofstadter1976,PhysRevLett.49.405,Wannier1978}. 

The unique spectral features of the HB have been a quintessential crystallization point for the 
integer quantum Hall effect \cite{Wannier1978,Koshino2001,Dean2013,Aidelsburger2015,Goldman_2009,Crosse2020,Koshino2002,Di_Colandrea_2022} and countless other many-body phenomena deriving from the intertwining of quantum dynamics, lattice regularization, and gauge invariance. Its discovery in physics had a significant interdisciplinary impact on mathematics, computer science, and popular culture \cite{GEBbook}. Mapping out the HB in any upcoming new geometry that is experimentally relevant is a sensitive and unique indicator for novel topological and dynamical phenomena that deserve closer experimental attention. In this Letter, motivated by groundbreaking experimental realizations of hyperbolic lattices in circuit quantum electrodynamics \cite{Kollar2019} and topolelectric circuits \cite{lenggenhager2021electric}, we study the Hofstadter problem on hyperbolic lattices \cite{Kollar2019,PhysRevX.10.011009,2019CMaPh.376.1909K,Boettcher2020,Yu2020,PhysRevD.103.094507,PhysRevD.102.034511,Maciejko2020,ZHANG20211967,Zhu2021,Boettcher2021,bienias2021circuit,lenggenhager2021electric,2021arXiv210602477L,maciejko2021automorphic,2021arXiv210905331M,2021arXiv210808854S}.

The physical magnetic fields required to achieve $\ell_B \sim a$ for typical lattice constants of solids are enormous ($B\sim10^5\text{T}$ for $a\sim 1$\AA{}), rendering it challenging to observe the HB directly in an experiment. Despite this obstacle, features of the HB have been observed in experiments with two-dimensional electron gases \cite{PhysRevLett.86.147}, optical lattices with large synthetic magnetic fields \cite{Struck2011,Aidelsburger2011,Miyake2013,Aidelsburger2013},  suspended graphene with a large effective lattice constant \cite{Dean2013}, and classical acoustic setups \cite{Ni2019}. The HB has also been explored in three dimensions, where its spectral properties depend on the direction of the magnetic field relative to the crystallographic planes \cite{Koshino2001,Koshino2003,Park2020}.

 To define the hyperbolic lattices relevant in this work, we say that a lattice is of $\{p,q\}$ type, with $p$ and $q$ positive integers, if it consists of $p$-gonal pla\-quettes with coordination number $q$ of each vertex, see Fig. \ref{Fig1}. A lattice of $\{p,q\}$ type is hyperbolic if $(p-2)(q-2)>4$, in which case it constitutes a regular tessellation of the hyperbolic plane. In contrast, for $(p-2)(q-2)=4$, the lattice is a regular tiling of the Euclidean plane; the only solutions are 
$\{4,4\}$, $\{6,3\}$, and $\{3,6\}$, corresponding to the square, honeycomb, and triangular lattices. 
The main finding of our work is that when the Hof\-stadter problem is studied for general $\{p,q\}$, a hitherto unnoticed universality comes to light: the shape of the butterfly is determined solely by the value of $p$. More precisely, the number of large extended gapped regions for $0\leq \Phi\leq \Phi_0/2$ is $p-2$. This universality of Hofstadter butterflies had been elusive just from studying the three regular Euclidean lattices, since they each exhibit different values of $p$. The universality is hinted at, however, in previous computations of the Hofstadter problem on two-dimensional quasicrystals with rectangular pla\-quettes \cite{Tran2015,Fuchs2016} and boundary mode spectra on hyperbolic $\{4,q\}$ lattices \cite{Yu2020}. In each of these cases, the shape of the $p=4$ square-lattice butterfly had been recovered.

\FigOne

On a lattice, magnetic flux is threaded through pla\-quettes by means of the Peierls substitution. When an electron travels along certain edges of the lattice, its wave function picks up a complex phase such that the total phase acquired when going around any plaquette is $e^{\rmi e \Phi/\hbar}$, see Fig. \ref{Fig1}. In order to observe the phenomena studied here in experimental electric circuit realizations of hyperbolic lattices, for instance, the Peierls substitution can be implemented by using phase elements that imprint a complex phase along edges connecting the nodes. Such phase elements have been developed recently \cite{AlbertaWu2021}. Crucially, the magnetic field strength can be tuned freely in this setup, allowing to reach the Hofstadter regime $\ell_B\sim a$ with $\Phi/\Phi_0\sim 1$.

The determination of the HB on hyperbolic $\{p,q\}$ lattices requires to overcome a substantial obstacle. Every finite planar subgraph of the infinite hyperbolic lattice has a macroscopic fraction of boundary sites. Further, going to larger subgraphs does not decrease this fraction \cite{Kollar2019,Boettcher2020}. As a result, the energy spectrum in a magnetic field is dominated by a significant contribution from boundary modes and thus does not feature the characteristic large extended gapped regions of the HB \cite{Yu2020,SOM}. The latter is defined as the pure bulk spectrum in the infinite system limit and so manifestly void of boundary contributions. We overcome this problem by a novel approach of studying the HB on compactified hyperbolic $\{p,q\}$ lattices which consist of up to several thousand vertices of equal coordination number $q$. These large but finite graphs are referred to as {\it regular maps} in the mathematical literature \cite{Conder2001,Conder2009, ConderWebsiteRegularMaps,Conder2002,Conder2006,ConderWebsiteTrivalentGraphs,ConderWebsiteGraphs30} and can be thought of as hyperbolic $\{p,q\}$ lattices with periodic boundary conditions that preserve all the local symmetries of the infinite lattice \cite{Coxeter1972,maciejko2021automorphic} an example of which is shown in  Fig. \ref{Fig1}.

The computation of the HB on a regular map is analogous to determining the spectrum of a finite Euclidean lattice with periodic boundary conditions on a torus, which is a surface of genus one. However, in the hyperbolic case, the Riemann surfaces that embed the regular maps have genus $g\geq 2$ and $g$ holes \cite{Maciejko2020}. The genus increases linearly with the number of vertices \cite{Boettcher2021}. To see this, we note that the genus of the embedding manifold is related to the Euler characteristic $\chi$ of the regular map via $\chi=2(1-g)$, where $\chi=V-E+F$ with $V$ the number of vertices (sites), $E$ the number of edges (nearest-neighbor bonds) and $F$ the number of faces (plaquettes) of the tiling. Regular maps additionally satisfy $pF=2E=qV$ \cite{Boettcher2021}, which proves that $g-1\propto V$. The infinitely many lattice sites limit is thus equivalent to $g\to \infty$. In our analysis of HBs on hyperbolic lattices, we verify that the HB for each $\{p,q\}$ is independent of $g$ (chosen sufficiently large, for a comparison see supplement Fig. S3 \cite{SOM}) and, therefore, a universal quantity determined solely by the values of $p$ and $q$. 

\FigTwo

Figure \ref{Fig2} shows the hyperbolic HB computed on regular maps of several representative $\{p,q\}$ tilings. In each case, the number of vertices of the regular map is chosen sufficiently large so that the HB can be considered converged to the infinite $\{p,q\}$ lattice spectrum. The first column contains the three Euclidean cases, whereas the other columns correspond to hyperbolic lattices. The coloring in the plots represents the normalized density of states, with darker colors indicating stronger degeneracy of the energy levels. Since the regular maps have coordination number $q$, their spectra lie in the real interval $[-q,q]$ for unit hopping amplitude. We observe that the overall shape of the HB solely depends on $p$. The number of extended gaps in the interval $ 0\le\Phi\le \Phi_0/2 $ is $ p-2 $. In the hyperbolic case, as $ q $ increases, the extended gaps in the spectrum become narrower. We verified the validity of these statements for additional choices of $\{p,q\}$ not included in the figure. The central result of this work is the calculation of the HBs from the bulk spectrum on finite hyperbolic lattices with periodic boundaries and the identification of these universal features.

In the following calculations, we use  $\hbar /e = 1$, then $\Phi_0=2\pi$. This choice of units eliminates the conversion factor between magnetic flux and hopping phases.
To compute the HB on a regular map, we realize a constant magnetic field by threading a flux $\Phi$ through each plaquette, as shown in Fig.~\ref{Fig1}. The particular assignment of complex phases to the bonds is not unique due to the presence of gauge degrees of freedom, which do not affect physical observables such as the energy spectrum. The perpendicular magnetic field imposes $F-1$ constraints on the $E$ hopping phases. The flux equation for the remaining plaquette is linearly dependent on the others and constrains $\Phi$ to rational values $\Phi/2\pi = n/F, \;n\in \mathbb{Z}$. Fixing the local $\text{U}(1)$ gauge eliminates another $V-1$ degrees of freedom in the hopping phases. This leaves 
$E -(V-1)-(F-1)= 2-\chi = 2g$ 
free parameters, which equals the number of Aharonov--Bohm (AB) fluxes that can be threaded through the holes of the manifold. The system of equations for the hopping phases is uniquely determined upon fixing the local gauge and the AB fluxes. The tight-binding Hamiltonian on the hyperbolic lattice is then encoded in the $V\times V$ Hermitian matrix $H_{p,q}(\Phi)$ whose $(i,j)$-entry is $e^{\rmi \phi}$ if sites $i,j\in V$ are connected by an edge with phase $e^{\rmi \phi}$, and zero otherwise. The exact diagonalization of this matrix yields the desired single-particle spectrum.

In the presence of a perpendicular magnetic field, there is no obvious choice of AB fluxes, since there is no point of reference for the contribution of the perpendicular magnetic field to the topologically non-trivial loops measuring the AB fluxes. However, all different possible configurations of AB fluxes can be mapped to actual eigenstates of the corresponding infinite $\{p,q\}$ lattice \cite{Maciejko2020,Boettcher2021}. We make use of this by averaging the calculated spectra over $\approx 10^3$ randomly sampled AB flux configurations, yielding greatly improved resolution in our numerical results. Plots of spectra for small genus or without AB flux sampling are shown in the supplement \cite{SOM}.

The most striking difference between Euclidean and hyperbolic HBs is the absence of a fractal structure in the non-Euclidean cases. This feature has not been unambiguously identified before because only our approach allows for ruling out boundary state contributions as a possible explanation for the missing fractality. The numerical observation can be understood partially from the modified Landau level structure in the hyperbolic plane. Indeed, in the Euclidean case, the discrete structure of the HB results from the highly degenerate Landau levels of the continuum model splitting into many bands that are no longer degenerate on the lattice; the extended gapped regions reflect the gaps between the original Landau levels. In contrast, as we explain in the following, the Landau level structure of the hyperbolic continuum model is different.

For the analysis of Landau levels, we consider the energy spectrum close to the lower band edge and for small $\Phi\ll~1$, where it can be approximated by a continuum model that ignores the discrete lattice. As has been shown in the seminal work of Comtet \cite{Comtet1985,Comtet1987}, the behaviour of a particle moving in the hyperbolic plane subject to a perpendicular magnetic field $B$ is substantially different from the Euclidean case. Due to the negative curvature that tends to divert trajectories, a sufficiently strong magnetic field is required to trap a classical particle onto a circular orbit. Similarly, in the quantum case, the energy spectrum of the continuum model features both a discrete and a continuous part \cite{Ludewig2021}. The discrete part consists of bound states or Landau levels with energies 
$\varepsilon_{n} = \kappa^{-2}[(2n+1)b-n(n+1)]$, with integers $ 0\le n \leq  n_{\rm max}<|b|-1/2 $, where parameter $b = \kappa^2 B$ is the rescaled magnetic field and $\kappa$ the curvature radius. Henceforth we set $\kappa=1$. The first term in the energy is reminiscent of the Euclidean expression, whereas the second term is a curvature-induced correction. The continuous part due to scattering states covers the energy range $\varepsilon_{\rm c}\geq \frac{1}{4}+b^2$. The mapping of the energy spectrum of the continuum model, $\vare_n$ and $\varepsilon_{\rm c}$, to the energies of the $\{3,7\}$ lattice model is shown in Fig. \ref{Fig3}  \cite{SOM}.

\FigThree

Another reminiscence of the continuum model in the lattice HBs is that the large extended gap regions become narrower as $p$ or $q$ increases. In the continuum model, the magnetic field required to trap particles onto a circular orbit increases with curvature, and the distance between separate Landau levels shrinks.  Since the enclosed curvature per plaquette, ${-p \pi (1 - 2/p - 2/q)}$, increases with $p$ and $q$, the narrowing of the large extended gap region is a remnant of the reduced Landau level spacing. Further, the effective magnetic field for fixed flux $\Phi$ decreases with increasing $p$ and $q$, because the area of plaquettes (proportional to the curvatures) increases.

It is instructive to compare our novel method for computing the HB as the bulk spectrum on graphs with periodic boundary conditions with the determination of topological boundary mode spectra on hyperbolic lattices with {\it open} boundary conditions in Ref. \cite{Yu2020}, where tessellations of type $\{4,q\}$ have been considered. The number of boundary sites approximately gives the number of boundary modes on a graph. Since the fraction of boundary sites converges to a finite, sizeable number in hyperbolic graphs with open boundaries, one cannot eliminate the boundary state contribution to the spectrum by merely considering large graphs \cite{SOM}. Instead, as suggested in Ref. \cite{Yu2020}, one may divide the eigenstates into bulk- and boundary-states through a suitable criterion, and estimate the bulk spectrum in this manner. We have performed this analysis for the hyperbolic lattices discussed in this work and found consistent results. However, the true form of the bulk spectra shown in Fig. \ref{Fig2} and questions about their fractality cannot be addressed by this approximate method.

Our results on the universality of the HB in $\{p,q\}$ lattices can be verified in electric-circuit realizations of hyperbolic lattices with open and periodic boundary conditions \cite{lenggenhager2021electric}.  Since the physical location of electric nodes in the network is irrelevant, any graph can be realized in principle. For hyperbolic graphs with open boundaries, which are planar, several thousand vertices are technically feasible; for graphs with periodic boundaries, the number is smaller in practice due to the need to connect edges across opposite sides of the graph, but a hundred sites could be possible in the near future. Different Euclidean lattices have been realized in this manner \cite{Lee2017,Helbig2019}, including networks with complex phases to simulate a nonzero magnetic field \cite{Hofmann2019}. A novel circuit element explicitly developed in the context of hyperbolic lattices implements a variable complex hopping phase and allows us to tune $\Phi$ experimentally to any value in the future \cite{AlbertaWu2021}. The possibility of measuring the spectrum site-resolved enables the separation of bulk from boundary states and determining the approximate bulk spectrum experimentally. In this manner, it is possible to verify the universal structure of HBs, namely that the number of large gaps in the bulk spectrum is determined by $p$. In addition, it is tantalizing to study the interplay between interactions experimentally due to nonlinear circuit elements and the magnetic
field to explore the fractional quantum hall effect in hyperbolic space. As a novel theoretical breakthrough, we realized large hyperbolic lattices with periodic boundary conditions that can be used to study quantum many-body systems and topological phases of matter in hyperbolic space in the future.

\begin{acknowledgments}
\textit{Acknowledgments:}
We gratefully acknowledge inspiring discussions with Anffany Chen, Joseph Maciejko, Canon Sun. AS, LKU and RT acknowledge support for funding from the Deutsche Forschungsgemeinschaft (DFG, German Research Foundation) through Project-ID 258499086 - SFB 1170 and through the W\"urzburg-Dresden Cluster of Excellence on Complexity and Topology in Quantum Matter -- \textit{ct.qmat} Project-ID 39085490 - EXC 2147.  IB acknowledges support from the University of Alberta startup fund UOFAB Startup Boettcher and Natural Sciences and Engineering Research Council of Canada (NSERC) Discovery Grants RGPIN-2021-02534 and DGECR2021-00043. 
\end{acknowledgments}

\onecolumngrid

\setcounter{equation}{0}
\renewcommand{\theequation}{S\arabic{equation}}
\setcounter{figure}{0}
\renewcommand{\thefigure}{S\arabic{figure}}

\section{S1. Continuum model}

\noindent In this section, we determine the relationship between the spectra of (i) the continuum model for a quantum particle moving on the hyperbolic plane and (ii) the lattice model of a quantum particle moving on a hyperbolic $\{p,q\}$ graph. We first assume absence of the magnetic field, setting $\Phi=0$, and will incorporate the magnetic field at the end of the section.

{\it Zero magnetic field ($\Phi=0$).} The continuum model Hamiltonian is (minus) the hyperbolic Laplacian
\begin{align}
	\Delta_g = \frac{1}{(2\kappa)^2}(1-|z|^2)^2(\partial_x^2+\partial_y^2),
\end{align}
acting on coordinates $z\in \mathbb{D}$ with hyperbolic Poincar\'{e} disk $\mathbb{D}=\{z\in \mathbb{C},\ |z|<1\}$.  The hyperbolic distance between two points $z,z'\in\mathbb{D}$ is 
\begin{align}
	d(z,z')=\kappa\ \text{arcosh}\Bigl(1+ \frac{2|z-z'|^2}{(1-|z|^2)(1-|z'|^2)}\Bigr).
\end{align}
In the following, we set the curvature radius $\kappa=1$. The lattice model Hamiltonian is given by (minus) the adjacency matrix of the finite hyperbolic graph, with entries being either $1$ or $0$. We denote the $V\times V$  adjacency matrix by $A$. The Schr\"{o}dinger equation for the lattice model is given by
\begin{align}
	-\sum_{j} A_{ij} f(z_j) = E f(z_i)
\end{align}
for all sites $z_i$. The sum runs over the sites $z_j$ of the graph. Assume the coordination number of site $z_i$ is $q$. This is guaranteed for regular maps, which have constant coordination number $q$ for each site. It has been show in Ref. \cite{Boettcher2020} that the latter equation can then be written as
\begin{align}
	\label{ig1} -\sum_{a=1}^q f\Bigl(\frac{z_i-w_a}{1-w_a\bar{z}_i}\Bigr) = E f(z_i),
\end{align}
with $w_a = h e^{2\pi(a-1)\rmi/q}e^{\rmi \chi_i}$, effective hyperbolic lattice constant
\begin{align}
	h = h(q,p) = \text{tanh}[d_0/(2\kappa)],\ d_0=d(r_0,r_0e^{2\pi \rmi/p}),\ r_0 =\sqrt{\frac{\cos(\frac{\pi}{p}+\frac{\pi}{q})}{\cos(\frac{\pi}{p}-\frac{\pi}{q})}},
\end{align}
and $e^{\rmi \chi_i}$ a phase that depends on $z_i$. Depending on the values of $p$ and $q$, the parameter $h$ may be small and can then be used for a perturbative expansion of Eq. (\ref{ig1}) in powers of $h$. The leading order term in this expansion is given by \cite{Boettcher2020}
\begin{align}
	-E f(z_i) &= q\Bigl [ f(z_i) + h^2 \Delta_g f(z_i) +\mathcal{O}(h^3)\Bigr].
\end{align}
Higher order terms contain both powers of $\Delta_g$ and more complicated differential operators. The first term that features the latter, more complicated differential operators is the term multiplying $h^q$ in the expansion of Eq. (\ref{ig1}). Terminating the series at a lower order in $h$ yields a polynomial in $\Delta_g$ that can easily be determined. In particular, for $q\geq 7$ we find
\begin{align}
	-E f(z_i) &= q\Biggl [ 1+ h^2\Delta_g + \frac{h^4}{4}(\Delta_g^2 +2\Delta_g)+\frac{h^6}{36}\Bigl(\Delta_g^3   +10\Delta_g^2 + 12 \Delta_g\Bigr) +\mathcal{O}(h^7)\Biggr]f(z_i).
\end{align}

{\it Nonzero magnetic field ($\Phi\neq 0$)}. We now estimate the eigenvalues $E$ for finite flux $\Phi>0$ by replacing $\Delta_g$ with the eigenvalue spectrum of a quantum particle moving in the hyperbolic plane with a perpendicular magnetic field $B$. As discussed in the main text, this spectrum consist of a discrete and continuous part, $\vare_n$ and $\vare_{\rm c}$ \cite{Comtet1985,Comtet1987}. We replace 
\begin{align}
	-\Delta_g \to \vare(b) &= \begin{cases} \frac{1}{4}+b^2+k^2 & (\text{continuum}, k\geq 0) \\ (2n+1) b - n(n+1) & (\text{Landau levels})\end{cases}\\
	&= \begin{cases} \frac{1}{4}+b^2+k^2 & (\text{continuum}, k\geq 0) \\ b\ \text{ for }b> 1/2 & (\text{lowest Landau level, }n=0) \\ (3 b - 2)\ \text{for }b>3/2 & (\text{second Landau level, }n=1) \\ \dots \end{cases}.
\end{align}
Parameter $b$ is the re-scaled magnetic field defined as $b=\kappa^2 B$.
Our estimate for the eigenvalue spectrum of the $\{3,7\}$ lattice model in a magnetic field close to the lower band edge becomes
\begin{align}
	E(b) = -7 \Biggl [ 1- h^2\vare + \frac{h^4}{4}(\vare^2 -2\vare)+\frac{h^6}{36}\Bigl(-\vare^3   +10\vare^2 - 12 \vare\Bigr) +\mathcal{O}(h^7)\Biggr],
\end{align}
with $h=0.496970$ for $\{p,q\}=\{3,7\}$. 
The value of the perpendicular magnetic field  $B$ is obtained from the quotient of flux $\Phi$ and the area per plaquette $A_p= -p \pi (1 - 2/p - 2/q)\kappa^2$, so $b=\frac{\Phi}{-p \pi (1 - 2/p - 2/q)}$. The resulting bands are shown in Fig. 3 of the main text.

\FigureSOne

\section{S2. Hofstadter Butterfly on open boundaries}

\noindent In this section, we explore the HB spectrum in the hyperbolic case with open boundary conditions. A comparison to the Euclidean case is also carried out.

The geometry of the lattice is analogous to Fig. 1 of the main text, however, with truncated boundaries, as shown in Fig. \ref{SFig1}$ (a) $. We consider a ring-like geometry that preserves a discrete rotation symmetry and where the number of sites in the ring progresses radially. For instance, the first ring (denoted by $ r_{1} $) contains the central $ p $-polygon, as shown in blue for $ \{3,7\} $. The second ring ($ r_{2} $, shown in red) contains all neighboring polygons of the central one. The constant magnetic field is induced by threading a flux $\Phi$ through each plaquette inside the ring. Unlike for periodic boundaries, there is no constraint on the allowed values of $ \Phi $. Increasing the number of rings $n$ results in a bigger system with denser energy spectrum. In the hyperbolic case, as $n$ increases, the fraction of boundary sites in comparison to the bulk sites (defined as sites contained in $ r_{n-1} $), $ N_{\text{bd}}/N_{\text{total}} $, saturates to a finite and significant value, \ref{SFig1}$ (b) $. In addition, due to exponential growth of the lattice for large $n$, we could only calculate using up to 12 rings ( $ r_{n = 12} $). In contrast, in the Euclidean case, the ratio $ N_{\text{bd}}/N_{\text{total}} $  approaches zero asymptotically. Thus, as $n\to \infty$ in the Euclidean case, only bulk states remain, and the spectrum approaches that of the HB, which receives no contributions from boundary states. In the hyperbolic case, the contribution from the boundary states does not diminish for open boundary conditions, but rather dominates over the bulk states.

Results for a selection of HBs on lattices with open boundary conditions are shown in Fig.~\ref{SFig2}. The top row displays the usual Euclidean cases, where the color in the plots signifies the localization of the states such that green color represents contributions from bulk states and red color those from boundary sites. We define bulk and boundary states $\Psi$ by calculating the position expectation value in the Poincar\'{e} disk given by  $\langle r\rangle =\sum_i r_i|\Psi(r_i)|^{2}$, assuming $\Psi$ normalized. We define bulk states as those with $\langle r\rangle \in [0,0.65]$ and boundary states as those with $\langle r\rangle \in [0.7,1]$.  This value also amounts to the color scheme mentioned before. The second row in Fig.~\ref{SFig2} displays the hyperbolic cases, where an enormous contribution from the boundary states is clearly visible in the spectra. In the lower part of the spectrum, there is an extended contribution from the bulk bands (green colored). This part is the contribution coming from the lowest Landau level, as discussed in the main text. For even $p$, when the lattice is bipartite, the upper part of the spectrum is obtained from the lower one by means of the particle-hole (sublattice) symmetry of the Hamiltonian.

\FigureSTwo

\section{S3. Calculating the Hofstadter Butterfly on regular maps}

\subsection{Calculating Hofstadter's butterfly}
\noindent Using the closed surface $\{p,q\}$ patterns (regular maps) described below, the HB spectrum can be computed from diagonalization of the $V \times V$ matrix $H_{p,q}(\Phi)$. We verify that the eigenvalues for fixed $\{p,q\}$ converge to a common limitting spectrum as the number of vertices (and thus the genus) increases. We interpret this limiting spectrum as the HB on the infinite $\{p,q\}$ lattice. In addition to working on large graphs, we make use of sampling over AB fluxes to obtain a spectrum with improved resolution. The convergence of both methods is exemplified for the $\{3,7\}$ lattice in Figs. \ref{Fig:A3} and \ref{Fig:A4}.

{\it Rational flux.} Since the sum of phases around any one face of the tiling is equal to minus the sum of the flux through all other faces, the total flux through all faces must always sum to zero for any choice of hopping phases. Nonetheless, a constant magnetic field can be applied to the lattice by using the fact that only fluxes modulo $2\pi$ are physically relevant, because phase factors $e^{\mathrm{i} \varphi_{ij}}$ are $2\pi$-periodic in the hopping phases $\varphi_{ij}$. This means that a constant magnetic field of flux $\Phi$ per face needs to fulfill
\begin{align}
	F \Phi = 2\pi n, \; n \in \mathbb{Z} \quad \iff \quad \Phi = 2\pi \frac{n}{F}.
\end{align}
The flux per face can only take on multiples of $2\pi$ divided by the number of faces $F$.

\FigureSThree

{\it Calculation procedure.} In order to calculate Hofstadter's Butterfly for regular $\{p,q\}$ tilings using a regular map of that tiling, we need (i) a graph of the regular map and its according adjacency matrix, (ii) a list of the face-cycles of the graph representing the faces of the regular map, and (iii) a set of cycles representing the AB loops of the map. 
For all following steps, we assume a fixed gauge is applied, for example based on a spanning tree.

We then calculate the spectrum of the hopping model for all allowed values of magnetic field \mbox{$\Phi = 2\pi n/F, \; n\in\{0,1,\ldots, F-1\}$}. For each value of $\Phi$, we sample a set of different AB fluxes, which means that, when going to the infinite plane, that the eigenstates calculated are not only those that transform trivially when traversing the hyperbolic Bravais lattice, but a sample of all eigenstates that transform under $U(1)$ between the Bravais lattice sites, which represents a larger subset of eigenstates of the hyperbolic plane. Even for a moderately sized graph, uniform sampling of the AB fluxes is a hopeless endeavor since the number of allowed values grows almost exponentially with the number of lattice sites, so we resort to random sampling. For a graph of $\approx 10^2 - 10^3$ sites, a similar amount of random samples is usually sufficient to obtain a reasonably low level of noise in the density of states, indicating that the calculation is converging to the exact result.

In the following, $\bm\varphi$ denotes an $E$-component vector ($E=$ number of edges) where each component is the phase of one edge in the hopping graph. It receives contributions from two effects: the external magnetic field and the AB fluxes. To determine the AB fluxes, we first calculate a basis of AB hopping phases $\{\bm{\varphi}^{\rm AB}_\mu\}$ by solving the system of equation for all fluxes equal to zero except for the $\mu^{\rm th}$ AB flux which is set to one.
The $2g$ degrees of freedom of the AB fluxes can be parameterized by a $2g$-component momentum $\textbf{k}$. The corresponding hopping phases are a linear combination of the AB phase basis $\bm{\varphi}^{\rm AB}(\bm{k})=\sum_{\mu=1}^{2g} k_{\mu}\, \bm{\varphi}^{\rm AB}_\mu $. For the constant magnetic field, we solve the system of equations for constant flux $\Phi=1$ and AB fluxes set to zero. This gives us a configuration of fluxes $\bm{\varphi}^{\rm m}$. For magnetic field $\Phi=2\pi n/F$, the hopping phases then are $\Phi\, \bm{\varphi}^{\rm m}$. In total, the hopping phases for some constant magnetic flux $\Phi$ and choice of AB fluxes specified by $\bm k$ is
\begin{align}
	\label{EqPhi} \bm\varphi(\Phi, \bm k) = \Phi \,\bm{\varphi}^{\rm m} +
	\sum_{\mu=1}^{2g} k_\mu \, \bm{\varphi}^{\rm AB}_\mu.
\end{align}
We insert these hopping phases into the adjacency matrix to obtain the matrix representation of the  hopping Hamiltonian $H_{p,q}(\Phi, \bm k)$ as a Hermitean $V\times V$ matrix. The spectrum of $H(\Phi, \bm k)$ is calculated for all allowed values of $\Phi$ and a random sampling of $\bm k$, and plotted over $\Phi$ to obtain the butterfly. The density of states as displayed in the various figures is calculated by binning the calculated spectrum, where the interval $[-q,q]$ is divided into $F$ equally spaced bins, and normalized by dividing by its mean value over said interval.

\subsection{Regular maps}

\noindent In order to determine the bulk spectrum of a hyperbolic $\{p,q\}$ tessellation without boundary contributions, we need to identify finite $q$-regular graphs build from $p$-gons. (A graph is called $q$-regular if the coordination number of every vertex is $q$.) These finite graphs cover closed surfaces. As far as we know, compactifying a given hyperbolic $\{p,q\}$ tiling to a closed cannot be accomplished by following a simple procedure as in the Euclidean cases, but rather constitutes a highly nontrivial combinatorical problem. Fortunately, extensive mathematical research on the classification of finite $q$-regular graphs provides us with most of the necessary input to define the underlying hyperbolic graphs for our calculations.

While our calculations can be performed on any $\{p,q\}$ tiling embedded on a closed manifold, it is worth discussing a particular class of them: regular maps. Regular maps, in addition to being $\{p,q\}$ tilings of closed surfaces, are required to possess the full rotational symmetry of the tiling, i.e. $p$-fold rotational symmetry around each face and $q$-fold rotational symmetry around each vertex. Regular maps are a convenient starting point because there exists a considerable amount of mathematical work on them. In particular, classifications are available in online databases, such as the one provided by Conder \cite{Conder2001, Conder2009, ConderWebsiteRegularMaps}. For many trivalent ($q=3$) or small regular maps, the adjacency matrices can be obtained by searching for regular maps in databases of symmetric graphs \cite{Conder2002, Conder2006, ConderWebsiteTrivalentGraphs, ConderWebsiteGraphs30}. 

In the following sections, we assume that the tilings we investigate are orientable and non-singular, meaning that no edges that belong to the same face are identified. The latter condition usually holds for regular maps with relatively small $p$ and $q$ and sufficiently large number of vertices. If it is violated, the correspondence between faces and cycles of the associated graph of the tiling fails, which complicates the identification of faces and incontractible loops.

\subsection{Counting free parameters and Euler's characteristic}
\noindent A closed, orientable surface is classified topologically by its integer-valued genus $g$, counting its number of holes. The genus of a closed manifold is related to the  Euler characteristic $\chi$ of a polyhedron embedded in it, by 
\begin{align}
	\chi = 2-2g. 
\end{align}
For a polyhedron with $V$ vertices, $E$ edges and $F$ faces, the Euler characteristic is defined as 
\begin{align}
	\chi=V-E+F. 
\end{align}
Now suppose we apply a magnetic field to the closed manifold. In the hopping model defined on the embedded polyhedral graph, this can be realized as a complex phase picked up along the edges (bonds) through to the Peierls substitution. The precise choice of phases is, however, dependent on the gauge of the vector potential of the magnetic field.

{\it Gauge transformations.} In the hopping model on the embedded graph, gauge transformations are represented by a diagonal unitary transformation $V\times V$-matrix, $\mathcal{U}$, attaching a complex phase factor to each vertex. Then the edge connecting two adjacent vertices picks up the difference of the gauge phase of the two vertices in its hopping phase. Since a global phase factor of $\mathcal{U}$ leaves all edges invariant, the number of gauge degrees of freedom in the choice of hopping phases is $V-1$. 
The gauge invariant quantities in the lattice are the fluxes that are threaded through the loops of the embedded graph. The flux through each face (plaquette) corresponds to the sum of the hopping phases along the encircling edges. The fluxes through the faces of the graph are interpreted as a magnetic flux flowing through the corresponding surface element of the closed manifold. Since, in a closed manifold, the oriented loop around any one face is the sum of all oriented loops around the other faces, specifying the magnetic field through the surface produces $F-1$ linearly independent equations.

{\it Aharonov--Bohm fluxes.}
Besides the fluxes through faces, there also exist fluxes threaded through or around the holes of the closed manifold, so-called Aharonov--Bohm fluxes (AB fluxes). The number of AB fluxes is equal to $2 g$, with $g$ the genus of the surface, since, for each hole, a magnetic flux loop can be threaded through it (in the space outside the closed manifold) or around it (in the space enclosed inside the closed manifold).

When we subtract the amount of independent equations and gauge degrees of freedom from the amount of free parameters (the $E$ hopping phases on the edges), we obtain
\begin{align}
	E-(V-1)-(F-1)-2g &=E-V-F+2-2g=0.
\end{align}
Thus, by fixing the magnetic field through the faces, the AB fluxes, and the gauge of the magnetic field, we exactly determine the complex hopping phase factor for each edge.

\subsection{Cycles and cycle spaces}
\noindent As described in the previous section, the flux through loops, either those enclosing faces or those around holes, represent the gauge invariant flux quantities for a closed surface in a magnetic field.

{\it Cycles.} A non-self-intersecting loop in a graph is called a \textit{cycle}. In a simple graph (all adjacency graphs fall under this), a cycle going from vertex $v_1$ to $v_2$, ect. to $v_n$ and then back to $v_1$ can be represented by the list of vertices it follows, $\{v_1, v_2, \ldots, v_n\}$. Note that there is no unique starting point and orientation to a cycle, so cyclic permutation or reversing the order of the vertices gives a list that describes the same cycle. If a cycle is equipped with an orientation, it becomes a \textit{directed cycle}, which can be represented the same way, but is only invariant under cyclic permutations of the vertices.

Another representation of cycles is as an edge vector. Suppose all edges of the graph are labeled according to $1,2,...,E$. Then a cycle can be represented by a vector with $E$ entries, where the entry is $1$ if the respective edge is part of the cycle and $0$ otherwise. In this representation, combining two cycles corresponds to addition of the edge vectors of the cycles modulo $2$. In this way, the cycles of a graph form a vector space that is a subspace of $\mathbb{Z}_2^E$. Linear operations on this vector space can be performed using mod-2 arithmetic.

{\it Contractible and incontractible loops.} In general, loops on a closed smooth manifold fall into two categories: \textit{contractible} loops that can be contracted to a point by continuous deformations, and \textit{incontractible} loops that cannot since they wind around holes in the surface. An important distinction between these two classes is that contractible loops enclose a surface on the closed manifold, i.e. there is a part of the manifold that can be designated as lying inside and another part as lying outside of the loop, with the inside section being simply connected.
On the embedded graph of the surface, cycles can be classified similarly in these two groups. For graphs, instead of continuous deformations, we consider adding or removing \textit{face-cycles} to the cycles. We consider a cycle contractible if it can be expressed as a linear combination of face-cycles and incontractible if not.

{\it Cycle basis.} A basis of the cycle space of a graph is called a \textit{cycle basis} of the graph. 
A cycle basis can be generated from any \textit{spanning tree} of the graph. For this, choose a particular spanning tree. It consists of $V-1$ edges. The remaining edges not contained in the spanning tree shall be denoted $\mathcal{E}$.  If we add an edge from $\mathcal{E}$ to the spanning tree, a cycle is formed. The set of cycles formed by adding each of the edges of $\mathcal{E}$ to the spanning tree are a cycle basis of the graph. This implies that the cycle basis has $E-V+1$ elements.

For us, a particular cycle basis is of particular interest: the one that consists of the independent faces of the graph and a set of linearly independent incontractible loops. It can be shown that this set of cycles is a cycle basis. Indeed, we know that there are $F-1$ linearly independent face cycles and $2g$ topologically inequivalent (and therefore linearly independent) incontractible AB cycles. By combining these two sets, we get a set of 
\begin{align}
	F-1+2g = F-1+(E-V-F-2)=E-V+1
\end{align}
linearly independent cycles, so they form a basis of cycle space. 

We can use this fact to construct an algorithm for finding the AB loops we need for our flux threading calculations. If we start with any cycle basis and find a subset of it that complements the face cycles to a complete cycle basis, then this subset is a set describing the linearly independent incontractible loops of the graph. This cycle basis can then be used to quantify the AB flux threaded through or around the holes of the surface. The linear dependence of a set of cycles, represented as edge vectors, can be tested by calculating the rank of their column matrix in mod-2 arithmetic.

{\it Gauge fixing.} In order to fix the gauge, $V-1$ constraints need to be imposed on the hopping phases in such a way that the set of total fluxes around the cycle basis remains linearly independent. A convenient way to achieve this is to fix the hopping phase of all edges that belong to a spanning tree of the graph. From the construction of the according cycle basis from the tree, it is clear that this leaves exactly one free flux parameter for every element of the cycle basis, so that it represents a valid choice of gauge.

\subsection{AB fluxes and the infinite hyperbolic plane}

{\it AB fluxes in the presence of magnetic fields.} In the absence of a magnetic field perpendicular to the plane, the AB fluxes of the closed manifold can be set to any desired value by equating the fluxes through the faces to zero and those through the AB loops to the desired values. Importantly, a continuous deformation, i.e. adding face cycles to the AB loops, does not affect the result since the flux through all faces is zero. However, this is no longer true when the magnetic field perpendicular to the plane is nonzero. Then a deformation of the AB loop will add the flux of the added face elements to the total flux picked up by the loop. There are essentially two ways to solve this problem: one either needs to find a particular arrangement of AB loops that correspond to the true value of AB flux, so that it can be fixed to zero, or perform an even sampling of AB fluxes, eliminating the effect of the offset introduced by the magnetic field. The former case is easy to solve in some lattices, such as the square lattice where straight paths form closed loops in the unit cell, but a general procedure appears to be much less trivial.

{\it Bloch states of the infinite lattice.} We find that the latter, alternative case of sampling AB fluxes with values in the range $[0,2\pi)$ has a physical motivation: States for different AB flux configurations represent Bloch wave states of the infinite hyperbolic lattice, and so sampling over the AB fluxes [or momenta $\textbf{k}$ in Eq. (\ref{EqPhi})] incorporates more states that would be present in a larger lattice whose unit cell is the regular map with $V$ vertices \cite{Maciejko2020,maciejko2021automorphic}. For this note that the AB fluxes of a $\{p,q\}$ regular map with genus $g$ can be used to calculate Bloch wave states of the infinite hyperbolic $\{p,q\}$ tiling. This is done by formally interpreting the $2g=n$ AB fluxes of a regular map as momenta $k_i \in \{k_1 \ldots k_n\}$ in an $n$-dimensional Euclidean lattice. This procedure is analogous to conventional two-dimensional band theory, where the unit cell can be thought of as a tiling of a torus (genus 1), and the two momenta correspond to the two AB fluxes of the surface. The regular map can then be used to tile the hyperbolic plane and the AB fluxes of the regular map provide a $U(1)$ cover of the resulting hyperbolic Bravais lattice. Bloch wave states are a subset of the eigenstates of the hyperbolic plane. Consequently, sampling over AB fluxes emulates larger lattices, which explains the convergence of the spectrum in Fig. \ref{Fig:A4} to a limit spectrum that agrees with the limit of a lattice with large number of sites (or genus) in Fig. \ref{Fig:A3}.

\end{document}